\documentclass[aps,prb,twocolumn,showpacs,groupedaddress,floatfix]{revtex4}  
\usepackage{graphicx}  
\usepackage{dcolumn}   
\usepackage{bm}        
\usepackage{amssymb}   
\usepackage{multirow}

\begin{document}

\title{Continuous-wave versus time-resolved measurements of Purcell-factors for quantum dots in semiconductor microcavities}

\author{M.~Munsch}
\author{A.~Mosset}
\author{A.~Auff\`eves}
\author{S.~Seidelin}
\email{signe.seidelin@grenoble.cnrs.fr}
\author{J. P.~Poizat}
\affiliation{CEA/CNRS/UJF Joint team ``Nanophysics and
semiconductors'',Institut N\'eel-CNRS, BP 166, 25, rue des Martyrs,
38042 Grenoble Cedex 9, France}
\author{J.-M. G\'erard}%
\affiliation{CEA/CNRS/UJF Joint team ``Nanophysics and
semiconductors'', CEA/INAC/SP2M, 17 rue des Martyrs, 38054 Grenoble,
France}

\author{A. Lema\^itre, I. Sagnes, P.~Senellart}
\affiliation{Laboratoire de Photonique et de Nanostructures,
LPN/CNRS, Route de Nozay, 91460 Marcoussis, France}

\date{\today}


\begin{abstract}
The light emission rate of a single quantum dot can be drastically
enhanced by embedding it in a resonant semiconductor microcavity.
This phenomenon is known as the Purcell effect, and the coupling
strength between emitter and cavity can be quantified by the Purcell
factor. The most natural way for probing the Purcell effect is a
time-resolved measurement. However, this approach is not always the
most convenient one, and alternative approaches based on a
continuous-wave measurement are often more appropriate. Various
signatures of the Purcell effect can indeed be observed using
continuous-wave measurements (increase of the pump rate needed to
saturate the quantum dot emission, enhancement of its emission rate
at saturation, change of its radiation pattern), signatures which
are encountered when a quantum dot is put on-resonance with the
cavity mode. All these observations potentially allow one to
estimate the Purcell factor. In this paper, we carry out these
different types of measurements for a single quantum dot in a pillar
microcavity and we compare their reliability. We include in the data
analysis the presence of independent, non-resonant emitters in the
microcavity environment, which are responsible for a part of the
observed fluorescence.

\end{abstract}

\pacs{42.50.Pq, 78.67.Hc, 78.90.+t} \maketitle

\section{Introduction}\label{intro}

Coupling an emitter to a cavity strongly modifies its radiative
properties, giving rise to the observation of cavity quantum
electrodynamics effects (CQED), which can be exploited in the field
of quantum information and fundamental tests of quantum mechanics. A
variety of systems allows one to implement different CQED schemes,
ranging from Rydberg atoms~\cite{Brune96} and alkaline atoms in
optical cavities~\cite{Rempe,Boozer}, to superconducting
devices~\cite{Blais}, as well for semi-conducting quantum dots (QDs)
(for an early review, see~\cite{JMtop}) coupled to optical
solid-state cavities. Thanks to impressive recent progress in
nanoscale fabrication techniques, vacuum Rabi
splitting~\cite{semicon,Hennessy}, giant optical non-linearities at
the single photon level~\cite{Englund07,rakher}, and vacuum Rabi
oscillation in the temporal domain~\cite{Englund} have been
demonstrated for single InAs/GaAs QDs coupled to microcavities.
Success in sophisticated CQED experiments requires first of all an
efficient enhancement of the spontaneous emission (SE) of an emitter
coupled to a resonant single mode cavity~\cite{Purcell}. The
dynamical role of the cavity is quantified by the so-called Purcell
factor $F$, namely the ratio between the emitter's SE rate with and
without the cavity. For an emitter perfectly coupled to the
cavity~\cite{perfectly} the Purcell-factor only depends on the
cavity parameters and takes on the value denoted $F_P$ which is
given by

\begin{equation}\label{FP}
F_P=\frac{3}{4\pi^2}\frac{Q}{V}\left(\frac{\lambda}{n}\right)^3.
\end{equation}
where $Q$ is the cavity quality factor, $V$ the cavity volume,
$\lambda$ the wavelength for the given transition, and $n$ the
refractive index.

The Purcell effect using QDs as emitters has first been observed
when coupled to pillar type microcavities in the late
90ies~\cite{Gerard98}. Moreover, when its radiation pattern is
directive, the cavity efficiently funnels the spontaneously emitted
photons in a single direction of space. This geometrical property
allows one to implement efficient sources of single
photons~\cite{Solomon,Moreau01,Press}, or even single,
indistinguishable photons~\cite{yamamoto,varoutsis}. A high Purcell
factor also enhances the visibility of CQED based signals like QD
induced reflexion~\cite{Englund07,Auffeves07}. Beyond its seminal
role, the Purcell factor appears thus as an important parameter
which measures the ability of a QD-cavity system to show CQED
effects, and has therefore become a figure of merit for quantifying
these effects. It is obviously important to develop reliable methods
to measure accurately this figure of merit.

Two types of measurements are possible. The first one is the most
intuitive and simply consists in comparing the lifetime of a QD at
and far from resonance with the cavity mode, using a time-resolved
setup~\cite{Gerard98}. This is feasible only as long as the resonant
QD lifetime is longer than the time resolution of the the detector,
or more generally longer than any other time scales involved, such
as the exciton creation time (capture and relaxation of electron and
holes inside the QD). For a large Purcell factor, this might be a
limiting condition. Instead, the Purcell effect can be estimated
from measurements under continuous-wave (CW)
excitation~\cite{bruno}. When approaching QD-cavity resonance, the
pump rate required to saturate the emission of the QD is higher due
to the shortening of the exciton lifetime. The Purcell effect also
produces a preferential funneling of the QD SE into the cavity mode,
and thus increases the photon collection efficiency in the output
cavity channel. Measuring either the saturation pump rate or the PL
intensity as a function of detuning, enables thereby one to measure
the Purcell factor.

This paper first aims at evaluating the consistency of these
different methods, and to compare their accuracy. Moreover, both
methods suffer from the same problem, related to the fact that the
cavity is illuminated by many other sources in addition to the
particular QD being studied. Even far detuned QDs can efficiently
emit photons at the cavity frequency. This feature has been observed
by several groups worldwide~\cite{Hennessy,Englund,Press,Kaniber},
leading to theoretical effort to understand this
phenomenon~\cite{Naesby,Yamaguchi,Kaniber,Auffeves}. All the models
involve the decoherence induced broadening of the QDs combined with
cavity filtering and enhancement. Even though one can easily isolate
the contribution of the single QD when it is far detuned from the
cavity mode, this becomes much more difficult near resonance when
other sources emitting {\it via} the cavity have to be taken into
account. With this aim, we have developed a model which includes
these contributions and therefore enables us to fit the experimental
data and to derive a correct value of the Purcell factor.

\section{Sample characteristics and setup}\label{sample}

\begin{figure}[t]{
\centering
\includegraphics[width=8.3cm]{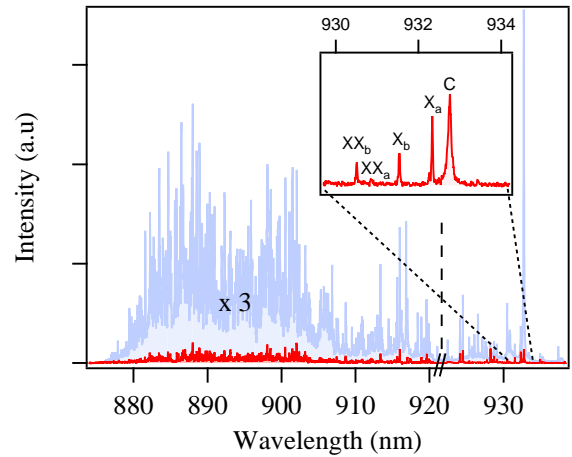}}
\caption{\label{raie} The full spectrum recorded at 4K showing the
inhomogeneous line, with a zoom on the section of interest including
two isolated quantum dots (X$_a$ and X$_b$) and their respective
bi-excitons (XX$_a$ and XX$_b$), and the cavity mode (C).}
\end{figure}

To fabricate the samples, a layer of InAs self-assembled QDs is
grown by molecular beam epitaxy and located at the center of a
$\lambda$-GaAs microcavity surrounded by two planar Bragg mirrors,
consisting of alternating layers of Al$_{0.1}$Ga$_{0.9}$As and
Al$_{0.95}$Ga$_{0.05}$As. The top (bottom) mirror contains 28 (32)
pairs of these layers. The quality factor of the planar cavity is
14000. In a subsequent step, the planar cavity is etched in order to
form a micro-pillar containing the QDs. The specific micro-pillar
discussed in the following has a diameter of 2.3~$\mu m$ and the
density of the quantum dots is approximately $2.5\times 10^{-9}$
QDs/$\rm cm^2$.

\begin{figure}[t]{
\centering
\includegraphics[width=8.3cm]{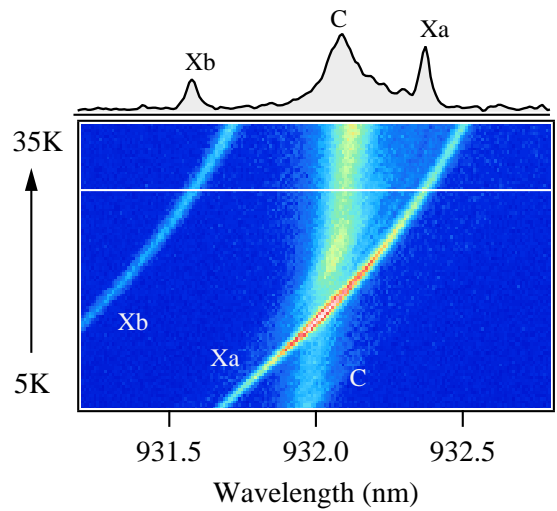}}
\caption{\label{temperature} PL spectra of cavity and QD when
varying the temperature (here from 5~K to 30~K). C indicates the
cavity mode, whereas X$_a$ and X$_b$ corresponds to the two QDs
spectrally closest to the cavity (see text). The white line
indicates the temperature for which the spectrum shown in the top
has been recorded.}
\end{figure}

The etching of the Bragg mirrors into a micro-pillar can deteriorate
the quality factor of the cavity. To measure the micro-pillar
quality factor, we perform a photoluminescence measurement at high
power such that the ensemble of QDs act as a spectrally broad light
source, which is used for probing the cavity~\cite{Moreau01,Gayral}.
From this measurement, we extract a quality factor of our specific
2.3~$\mu m$ diameter sample mentioned above of
$Q=\lambda/\Delta\lambda=7500$. This value agrees (to within 10\%)
with reflectivity measurement using white light. We will, in the
following section, use the corresponding bare cavity linewidth
$\kappa_0=\lambda/Q$ (in nanometers). Using equation~\ref{FP}
together with the measured value for the quality factor, we obtain
$F_P=18.6$.

Our sample is located in a cryostat held at 4K. For the
continuous-wave measurements, the QDs are excited using a standard
laser diode emitting at 820 nm (off-resonant excitation in the GaAs
barrier), while for the time resolved measurements, we use a pulsed
Ti:Sa laser centered at 825 nm (80 MHz repetition rate and 1 ps
pulse width). The emitted light is recollected after passing a
spectrometer (1.5 m focal and 0.03 nm resolution). The spectrometer
has two output channels: one channel leads to a CCD camera (for the
CW measurements), the other to an avalanche photo diode (APD) with a
40 ps time resolution which, combined with a 5 ps resolution for the
data acquisition card and 65 ps resolution due to the spectrometer,
gives us an overall resolution of 80 ps.

In fig.~\ref{raie} we give an overview of the different lines
observed in a typical photoluminescence experiment for our
particular micro-pillar to be studied in the following. Centered
around 895 nm, we observe what is usually referred to as the
inhomogeneous line, composed of hundreds of QDs. The micro-pillar
has been processed such that the cavity resonance is located on the
low energy wing of this inhomogeneous line, where the QD density is
very low, allowing us to optically isolate one single QD (denoted
$X_a$) to be studied, and in particular scanned through cavity
resonance. We also note that its corresponding bi-exciton ($XX_a$)
is blue shifted by about 1~nm, an amount which is larger than the
cavity linewidth. For a given temperature, we can therefore make the
bi-exciton off resonance with the cavity, while having the exciton
centered at resonance. For this specific micro-pillar, this happens
at 19.5~K. In this case, a second QD ($X_b$) appears about 3 cavity
linewidths away (with its bi-exciton $XX_b$ even further away), and
is therefore also minimally affected by the cavity. All other QDs
are much further detuned.

In fig.~\ref{temperature} we show the temperature dependence of the
cavity resonance frequency, as well as the two relevant QDs emission
wavelengths. The cavity frequency varies due to a temperature
dependent refractive index, while the QD exciton energy follows the
expected temperature dependence of the GaAs bandgap. Due to this
difference in temperature dependence, we can vary the cavity - QD
detuning~\cite{kiraz,semicon,Hennessy}.

\section{Continuous-wave measurements}

\begin{figure}[t]{
\centering
\includegraphics[width=8.3cm]{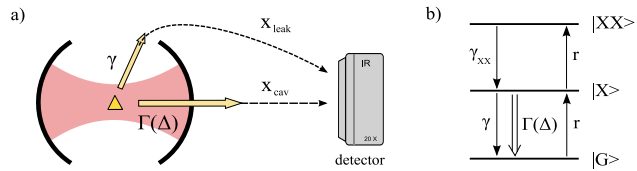}}
\caption{\label{cavite2} a) The PL of the QD arriving at the
detector can be separated into two channels: one part emitted into
loss channels ($\gamma$) but redirected to the detector with a
probability $\chi_{leak}$ and the part emitted into the cavity
$\Gamma(\Delta)$, and detected with a probability $\chi_{cav}$. b)
Three-level scheme including the exciton $\mid X\rangle$ and
bi-exciton $\mid XX\rangle$. The notations are defined in the text.}
\end{figure}

Even though the Purcell effect is a dynamical phenomenon, it can be
measured without a time resolved setup. This can be understood as
follows. As the emitter's lifetime decreases near resonance due to
the Purcell effect, it becomes harder to saturate the optical
transition. This can be quantified by measuring the increase in the
pump rate required to saturate the emitter (see
section~\ref{secpumprate}), or by measuring the actual cycling rate
in a PL measurement at saturation (section~\ref{secPLinten}). So by
comparing the on- and off-resonant saturation pump rate or PL
intensity, the Purcell-factor can be measured. More recently it has
been demonstrated that one can also extract the Purcell-factor due
to the change in the fraction of SE that is funneled into the cavity
mode~\cite{Gayral}. This is done by measuring the SE rate as a
function of detuning for fixed pump power, as will be done in
section~\ref{secfixedpumrate}.

\begin{figure}[t]{
\centering
\includegraphics[width=8.3cm]{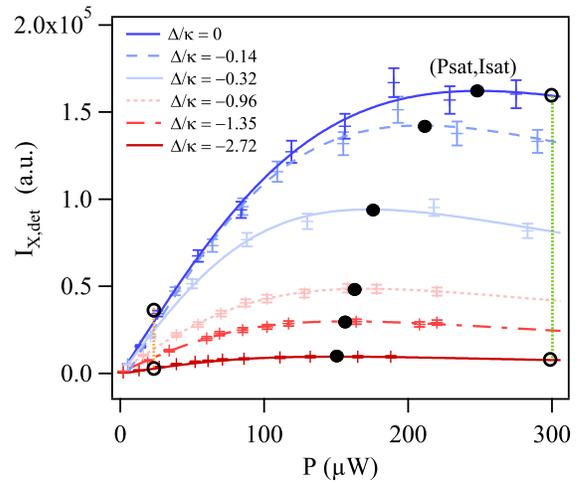}}
\caption{\label{Isat} Photoluminescence intensity (I) as a function
of pump power (P) for different cavity versus quantum dot detunings.
The open circles allows us to extract $\epsilon_{above}$ and
$\epsilon_{below}$ as described in section~\ref{secepsilon}. Filled
black circles indicate the saturation intensity I$_{sat}$, and the
corresponding pump power needed to saturate the QD, denoted
P$_{sat}$.}
\end{figure}

An illustration of the principles is given in fig.~\ref{cavite2}. A
QD is embedded in a cavity whose fundamental mode is nearly resonant
with the excitonic $X_a$ transition (fig. 1a). We denote $\Delta$
the detuning between the excitonic transition and the cavity mode.
The QD is non-resonantly pumped with a rate $r$, and decays by
emitting photons either in the cavity mode, or in other leaky modes
with a rate which we suppose to be independent of the detuning
$\Delta$ and identical to that of the bulk material (which is a
reasonable approximation for QDs in micro-pillar
cavities~\cite{Gerard98} and microdisks). As suggested by the PL
spectra shown in part II, the QD should be modeled by a three-level
system which includes the bi-exciton (fig.~\ref{cavite2}b). In the
following we will concentrate solely on $X_a$, so for simplicity we
will omit the subscript $a$. We denote $\gamma$ and $\gamma_{XX}$
the coupling of the X and bi-excitonic (XX) transition with the
leaky modes. In addition to the leaky modes, the X transition is
coupled to the cavity mode with a rate $\Gamma (\Delta) = \gamma F
{\cal L} (\Delta)$ where $F$ is the effective Purcell factor
experienced by the QD, taking into account that it is not perfectly
coupled to the cavity (in contrast to $F_P$ given in
equation~\ref{FP}, which is only an upper bound for $F$). Moreover,
${\cal L} (\Delta)=1/(1+\Delta^2/\kappa_0^2)$ is a Lorentzian of
width $\kappa_0$ corresponding to the empty cavity line shape. When
pumping with a rate $r$, the average excitonic population is then
given by

\begin{figure}[t]{
\centering
\includegraphics[width=8.3cm]{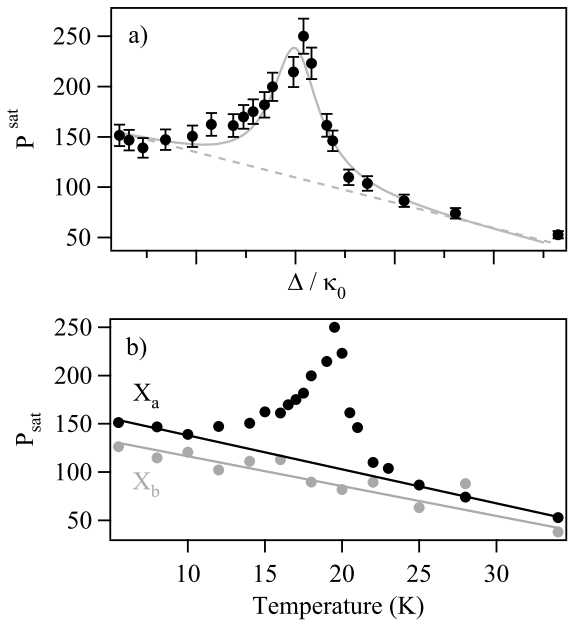}}
\caption{\label{sat} a) Saturation pump power as a function of
detuning for our single QD and b) saturation pump power for our
single QD $X_a$ and a ``control'' QD $X_b$ as a function of
temperature. The QD $X_a$ goes through the cavity resonance while
$X_b$ remains detuned throughout the scan.}
\end{figure}

\begin{equation}\label{pop}
p_X(\Delta,r)=\frac{1}{1+\frac{r}{\gamma_{XX}}+\frac{\gamma+\Gamma(\Delta)}{r}}.
\end{equation}
As it was mentioned in the first part of this paper, the role of the
cavity is not only to enhance the cycling rate for the exciton
($X$), but also to efficiently funnel the emitted photons into the
cavity mode. Provided its emission pattern is directional, which is
the case for micropillars, the coupling with a conveniently
positioned detector can be very efficient, whereas the coupling
between leaky modes and detector remains poor. These geometrical
efficiencies are respectively denoted $\chi_{cav}$ and $\chi_{leak}$
(see fig.~\ref{cavite2} and ref~\cite{Auffeves}). The PL intensity
from our single QD collected by the detector can thus be written in
the following way:

\begin{equation}\label{ID}
I_{X,det}(\Delta,r)=I_{X}^{leak}(\Delta,r)+I_{X}^{cav}(\Delta,r),
\end{equation}
where

\begin{equation}
I_{X}^{leak}(\Delta,r)= \chi_{leak} \gamma p_X(\Delta,r)
\end{equation}
is the PL intensity emitted through the leaky modes and

\begin{equation}
I_{X}^{cav}(\Delta,r)= \chi_{cav} \Gamma(\Delta) p_X(\Delta,r)
\end{equation}
the detected PL intensity emitted spatially into the cavity mode.
Please note that the notation $cav$ applies to geometrical
considerations, but not to the emission frequency (this PL
contribution is indeed emitted at the QD frequency). In our
experiment, to separate $I_{X,det}$ from the PL intensity from all
other light sources, we use of the spectrometer to focus on a window
centered on our selected QD (see the inset in fig.~\ref{raie}) and
we then fit the line shape corresponding to the single QD with a
Lorentzian function. When the QD-cavity detuning is large, it is
easy to separate the QD line shape from the cavity, but as the
detuning decreases, they will partially overlap with each other.
When this happens, to avoid a part of the cavity peak erroneously
being included in the single QD line shape, we also do a Lorentzian
fit on the cavity profile, which we then subtract from the QD line
shape. Note that in doing this, we also involuntarily omit from
$I_{X,det}$ the part of the QD PL which is emitted at the cavity
frequency, but this part constitutes a small fraction of the total
signal.

An example of typical experimental data is pictured in
fig.~\ref{Isat}, where the PL intensities for different detunings
$\Delta$ are plotted. As we generally measure the pump power denoted
$P$ and not the pump rate $r$, we have chosen to plot the data as a
function of the former (and we do the same in the graphs to follow).
This also means that $P^{sat}$ is the pump power corresponding to
the pump rate $r^{sat}$.

For each detuning, the maximal intensity
$I_{X,det}^{sat}(\Delta,r^{sat}(\Delta))$ is reached when the X
transition is saturated, where $r^{sat}(\Delta)$ denote the pump
rate required to saturate the transition (saturation pump rate).
Note that the highest values of $I_{X,det}^{sat}$ and corresponding
$r^{sat}$ are reached at resonance, which is coherent with the
enhancement of the X transition rate induced by the cavity. In the
following, we will analyze the curves presented in fig.~\ref{Isat}
(and further equivalent curves not added to the graph for clarity),
in four different ways (section~\ref{secpumprate}-\ref{secepsilon}).

\subsection{Saturation pump rate measurements}\label{secpumprate}

In the first method the Purcell factor is extracted from the
dependence of the saturating pumping rate intensity with the
detuning $r^{sat}(\Delta)$, corresponding to black filled circles in
fig.~\ref{Isat}. This method has been proposed as a substitute for
the time-resolved measurements, and has been widely used for
micropillars~\cite{bruno}, microdiscs~\cite{bruno} and photonic
crystals~\cite{happ}. The analytic expressions can be found by
determining the pump rate corresponding to the maximum intensity of
equation~\ref{ID}. We obtain

\begin{equation}\label{Rsat}
r^{sat}(\Delta)\propto \sqrt{1+F{\cal L}(\Delta)}.
\end{equation}

In fig.~\ref{sat}a we have plotted the data and the fit according to
equation~\ref{Rsat} where we have imposed the bare cavity linewidth
based on independent measurements. From the first fit, we extract a
Purcell-factor of

\begin{equation}
F=3.7\pm 1.0,
\end{equation}
where the relatively large error is due to the uncertainty of
$r^{sat}$. The slope of the baseline in fig.~\ref{sat}a is due to
the increase in temperature for increased detuning. As mentioned in
section~\ref{sample}, we use an optical excitation obtained through
the pumping of the GaAs barrier material. The mean free path of the
electrons and holes increase with temperature, so that the
excitation rate of the QD tends to increase for a fixed pump rate.
As a test, we have checked that the PL of another far detuned QD,
$X_b$, gives rise to an equivalent slope during the same experiment,
see fig.~\ref{sat}b.

\begin{figure}[t]{
\centering
\includegraphics[width=8.3cm]{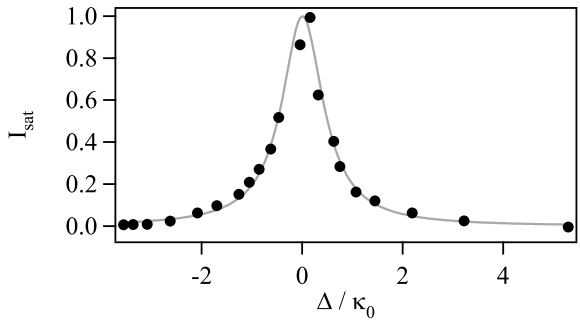}}
\caption{\label{Psat_T} Saturation intensity for the single QD as a
function of detuning. The size of the error-bars corresponds
approximately to the extent of the data points and are therefore not
shown.}
\end{figure}

\subsection{Saturation PL intensity measurements}\label{secPLinten}

Another similar approach again based on the black filled circles in
fig.~\ref{Isat} has been used in recent papers
~\cite{Bockler,pascale}. This method corresponds to exploiting
directly the maximum intensity of equation~\ref{ID} given by

\begin{equation}\label{Idsat}
I_{X,det}^{sat}(\Delta)\propto \frac{F{\cal L}(\Delta)}{1+\sqrt{2 +
2F{\cal L}(\Delta)}},
\end{equation}

In fig.~\ref{Psat_T} we have plotted the data and the fit (the
maximum normalized to one) according to equation~\ref{Idsat}, again
with the bare cavity linewidth fixed. From the fit, we extract a
Purcell-factor of

\begin{equation}
F=2.4\pm 1.2.
\end{equation}
In this case, the intrinsic uncertainty in the PL measurement is
quite small, but is amplified by the fitting procedure, resulting in
the stated errorbar.

\subsection{PL intensity with fixed pump rate}\label{secfixedpumrate}

\begin{figure}[t]{
\centering
\includegraphics[width=8.3cm]{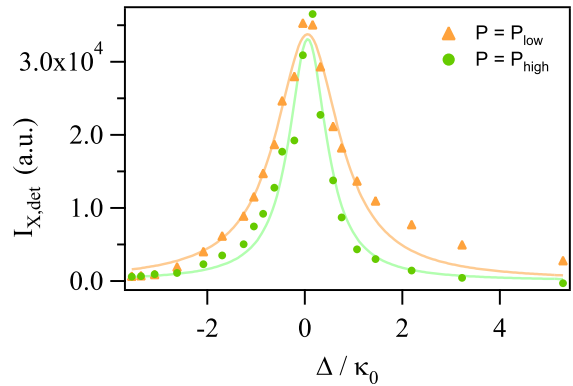}}
\caption{\label{Cloche} Measurements of the PL intensity at fixed
pump power (30~$\mu W$ and 300~$\mu W$, respectively). The two
curves can be thought of as the intersection of the curves in
fig.~\ref{Isat} with vertical lines centered at P=30~$\mu W$ and
P=300~$\mu W$ (with several more similar curves added).}
\end{figure}

In the two previous sections, we have used the data corresponding to
the saturation pump rate and intensity. Instead, we can also use the
emitted PL intensity, not at saturation, but for a fixed pump
rate~\cite{Gayral}. This amounts to using the PL intensity
corresponding to the intersection of the curves in fig.~\ref{Isat}
with a straight vertical cut. In particular, we have plotted the PL
intensity for powers below and above saturation in
fig.~\ref{Cloche}. The fit corresponds again to equation~\ref{ID},
but this time with the pump rate fixed ($r=30 \mu W$ and $r=300 \mu
W$, for the two curves, respectively). From both curves we have
subtracted a global offset corresponding to the PL intensity
$I_{X,det}$ at $\Delta=\infty$.

Below saturation the change in the light intensity $I_X^{cav}$ as
the QD is scanned across the cavity resonance is due to the
geometrical redirection of the emission alone (a modification in the
emission pattern). What we detect is a projection of a fraction of
the micro-pillar emission pattern onto the microscope aperture. More
precisely, for low powers (well below saturation)
$p_X(\Delta,r)=\frac{r}{\gamma + \Gamma(\Delta)}$, and we obtain

\begin{equation}
I_X^{cav}(\Delta)\propto \frac{ F{\cal L}(\Delta)}{1+F{\cal
L}(\Delta)}\equiv\beta(\Delta),
\end{equation}
where we have defined the function $\beta(\Delta)$ which can be
interpreted as the fraction of the emission pattern overlapping with
the cavity mode. This function is broader than the Lorentzian
profile of the cavity mode by a factor $\sqrt{(F+1)}$.

Above saturation, the geometrical redirection of the emission
pattern is still present but the light intensity $I_X^{cav}$ follows
now the ${\cal L}(\Delta)$ profile of the cavity owing to the
additional effect of the larger emission rate of the quantum dot
caused by the shortening of its lifetime. More precisely, in the
regime well above saturation we have
$p_X(\Delta,r)\approx\gamma_{XX}/r$, and we get

\begin{equation}I_X^{cav}(\Delta,r)\propto F{\cal L}(\Delta).
\end{equation}
From the ratio of the two widths, we extract a Purcell-factor of

\begin{equation}
F=3.2\pm 0.9,
\end{equation}
where the stated uncertainty arises from the intensity measurements,
which is the dominant source of error in this case.

\subsection{PL intensity ratio at low and high pump rate}\label{secepsilon}

This method also consists in comparing the light emitted by the
single QD at resonance and far from resonance, below and above
saturation, but only requires four of the measurements used above.
Here we do not subtract the offset due to $\chi_{leak}$ as done
above, which has the advantage that it allows us to quantify
$\chi_{cav}/\chi_{leak}$. We define as $\epsilon(\Delta,r)$ the
following ratio

\begin{equation}
\epsilon(\Delta,r)=\frac{I_{X,det}(0,r)}{I_{X,det}(\Delta,r)}=\frac{
p_X(0,r)}{p_X( \Delta, r )}\times
\frac{\chi_{leak}+\chi_{cav}F}{\chi_{leak}+\chi_{cav}F{\cal L} (
\Delta )}
\end{equation}
\begin{equation}
\equiv \frac{ p_X(0,r)}{p_X(\Delta,r)}\alpha(\Delta),
\end{equation}
where the parameter $\alpha(\Delta)$ depends on the cavity funneling
properties.

For pump rates below the pump rate required to saturate (where
$p_X(\Delta,r)=\frac{r}{\gamma + \Gamma(\Delta)}$)

\begin{equation}
\epsilon_{below}(\Delta)=\alpha(\Delta)\times\frac{1+F{\cal L}(
\Delta )}{1+F},
\end{equation}
whereas above the saturation pump rate (again using that
$p_X(\Delta,r)\approx\gamma_{XX}/r$)

\begin{equation}
\epsilon_{above}(\Delta)=\alpha(\Delta).
\end{equation}
Taking the ratio between $\epsilon_{below}$ and $\epsilon_{above}$,
$\alpha(\Delta)$ cancels, and  with an independent measurement of
$\kappa_0$ (see section~\ref{sample}), we obtain a Purcell-Factor of
\begin{equation}
F=2.5\pm 0.5,
\end{equation}

where the error arises from the uncertainty on the intensity
measurements. From the separate value of $\epsilon_{above}$ (or
$\epsilon_{below})$ we get

\begin{equation}
\chi_{cav}/\chi_{leak}\sim 15\pm 4.5,
\end{equation}

confirming that the cavity is much better coupled to the detector
than the leaky modes. This ratio depends on the radiation pattern of
the micro-pillar, and the numerical aperture of the collection
objective (0.4 for the above stated ratio of
$\chi_{cav}/\chi_{leak}$).

Note that we have only included the presence of exciton and
bi-exciton in all given formulas. We have, however, repeated the
above analysis, allowing for all orders of exciton levels, without
any significant change in final results within the range of used
pump powers.

\section{Time-resolved measurements}

\begin{figure}[t]{
\centering
\includegraphics[width=8.3cm]{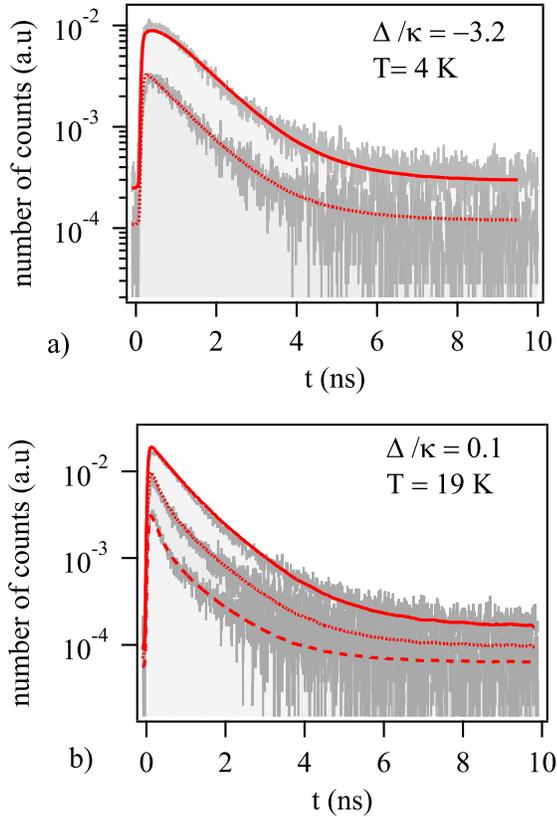}}
\caption{\label{TDV} Lifetime measurements at different pump powers
of the quantum dot while a) far detuned from the cavity and b) close
to resonance. In a) the solid line corresponds to $P=3P_{sat}$ and
dotted line to $P=P_{sat}$. In b) we have $P=P_{sat}$ (solid line)
and $P=P_{sat}/10$ (dotted line) and $P=P_{sat}/30$ (dashed line).}
\end{figure}

\begin{figure}[t]{
\centering
\includegraphics[width=8.3cm]{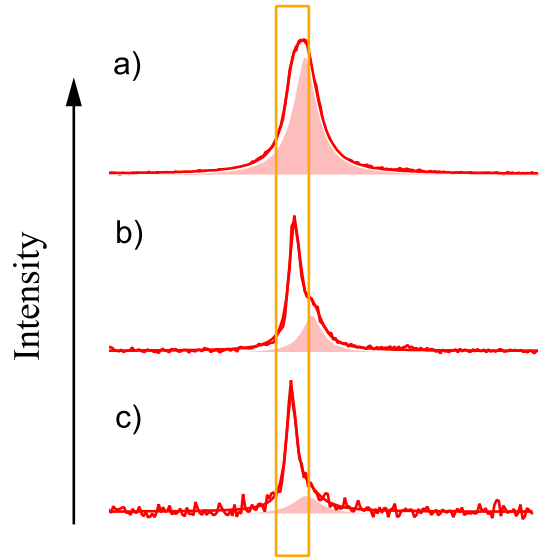}}
\caption{\label{spectres} Three different line spectra, each
corresponding to the QD (solid line) and the cavity (shaded area).
The spectra are shown for the pump power decreasing from a) through
c). For all three cases, the detuning is fixed ($\Delta=-0.2
\kappa_0$). The square frame indicates for each spectrum the
integration window.}
\end{figure}

As a way to confirm our continuous-wave measurements of the Purcell
factor, we have performed a detailed study of the lifetime as a
function of the detuning, using time resolved spectroscopy. This
technique has been used extensively for many different systems since
it was the first method to be used. In fact, the Purcell factor can
be written as

\begin{equation}
F= \frac{\tau(\Delta=0)}{\tau(\Delta=\infty)}-1,
\end{equation}
where $\tau$ is the lifetime of the QD, and $\Delta$ again is the
detuning. Opposite equation~\ref{FP}, this definition also applies
to an emitter that is not perfectly coupled to the cavity (within
the approximation where $\gamma_{leak} = \gamma_{bulk}$, the latter
denoting the SE of the QD into the unprocessed, or bulk, material).

In fig.~\ref{TDV} we show the measured lifetime of our quantum dot
for different pumping powers. In a) the QD is detuned from the
cavity resonance, while in b) it is at resonance. In the first case
(a), we show two different powers. When $P\leq P_{sat}$ (lowest
lying curve) the QD exhibits the typical mono-exponential decay.
When $P\geq P_{sat}$ (highest lying curve), the effect of the
bi-exciton can be observed as a rounding off of the curve at short
time, which corresponds to the delay in the radiation of the
exciton. The data fit very well with a model including three levels
(a ground state, the exciton and bi-exciton states) and we extract
the exciton and bi-exciton lifetimes, which are the same for the two
different powers:
\begin{equation}
 \tau_X= 0.80\pm 0.05~{\rm ns}~~~{\rm and}~~~~
\tau_{XX}= 0.40\pm 0.02~{\rm ns}
\end{equation}
As the biexciton is not influenced by the Purcell effect (for the
detunings used in this experiment), the obtained value can be used
as a fixed parameter when we then fit the data for the resonant
case. Note that all our fits have been convoluted with the
experimental system response time (80 ps time resolution). On the
contrary, the resonant case (b), shows a power dependency that can
not be explained by our simple three level model used above. We
clearly observe in fig.~\ref{TDV}b a change from a quasi
mono-exponential decay to a bi-exponential decay, when lowering the
pump rate. We exclude a prominent role of dark exciton since a
mono-exponential behavior is observed in the non-resonant case (a).
In addition, the fact that the second lifetime of the exponential
decay is fast (less than 1~ns) also tends to eliminate this
hypothesis. We believe that this behavior is due to detuned
emitters, which contribute to the collected intensity via the cavity
emission. Recent experiments~\cite{Hennessy,Englund,Press,Kaniber}
show that QD's could emit photons in the cavity mode even at rather
high detunings (several times the cavity linewidth). In contrast to
CW measurements where we could separate the emission of our QD from
the one of the cavity using appropriate lorentzian fits, in the
present case we do not have access to the full spectra, and
therefore cannot use the same technique. Instead we must select a
frequency window around the QD line, for which we integrate all PL.
This makes us unable to filter out the cavity component which
overlaps in frequency with the chosen window (when close to
resonance, as in fig.~\ref{TDV}b). As a result we measure two
different times: the shorter one is the lifetime of our single QD
(undergoing Purcell effect), whereas the longer one corresponds to
the lifetime of other detuned emitters. The higher the pump power,
the more dominant is the signal due to the contribution of the
detuned emitters. Therefore, at high powers, the light from other
emitters tends to make the signal invisible for our single QD. This
is illustrated in fig.~\ref{spectres}, where we have shown the
spectra corresponding to three different pump powers, ranging from
high (a) to low (c), but for a fixed detuning. The fraction of light
emitted {\it via} the cavity clearly dominates at high powers, but
decrease while lowering the pump power.

This is why, for high powers, only one lifetime can be observed
(upper curves in fig.~\ref{TDV}b), and this lifetime is obviously no
longer the QD radiative lifetime, but corresponds to the light
emitted {\it via} the cavity. Only for lower pump power, the true
lifetime also becomes visible (lower curve) as seen by the
bi-exponential decay. We therefore need to include these additional
emitters that we can model (within our pumping range) with a
two-level system whose lifetime corresponds to an average lifetime,
which can be measured in an independent experiment, in which all QDs
are far detuned. We obtain $0.8\pm 0.05$~ns.

\begin{figure}[t]{
\centering
\includegraphics[width=8.3cm]{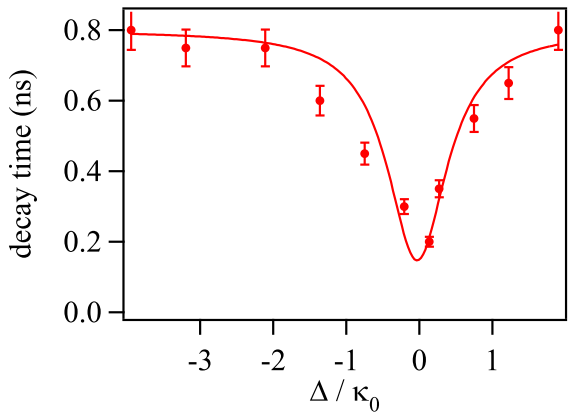}}
\caption{\label{TDV_DELTA} Exciton lifetimes measured at low
intensity as a function of QD-cavity detuning.}
\end{figure}

The exciton lifetime is the only free parameter in our fits (the
bi-exciton lifetime is a fixed parameter). The excellent agreement
between data and fit seems to validate our model. We find for the
non resonant and resonant case, $\tau(\Delta=\infty)=0.80 \pm
0.05$~ns and $\tau(\Delta=0) =0.2 \pm 0.01$~ns, which gives a
Purcell factor of:
\begin{equation}
F = 3.0\pm0.5
\end{equation}
where the stated uncertainty arises from the exponential fit. Based
on the above discussion, we remark that the low power condition is a
necessary but not sufficient criterium for measuring the correct
lifetime. Indeed, though all shown powers in fig.~\ref{TDV} are
below saturation only the complete model gives the right lifetime.
In fig.~\ref{TDV_DELTA} we have plotted the exciton lifetime
obtained by measurements equivalent to those in fig.~\ref{TDV} for
different values of the detuning. The shape of the curve should be
the Lorentzian profile of the cavity, confirmed by the fit.

\section{Final discussion}

We have presented several ways to measure the Purcell-factor, which
is an important figure of merit in cavity QED. All our CW
measurements agree with each other, within the experimental
uncertainty, for a Purcell-factor of $3.0\pm 0.4$. We emphasize that
in our evaluation of the error-bars, we have not taken into account
the stated 10 \% uncertainty for the bare cavity linewidth (see
section~\ref{sample}). A simple PL measurement of the cavity
linewidth has negligible error-bars, but when probing the cavity by
reflectivity measurements, this value turns out to be about 10 \%
different. We also point out that the value measured by reflectivity
is systematically higher than the one measured in PL. We will here
revisit the obtained results for the Purcell factor, in order to see
how a 10\% deviation on the quality factor would affect the values.
While the first method (based on saturation pump power, in
section~\ref{secpumprate}) does not depend on this parameter, all
the other CW methods here presented do. In particular, the second
technique, which uses the saturation intensity (\ref{secPLinten}),
drastically depends on this parameter. In our case, an uncertainty
of 10\% on the quality factor would make the measurement based on
this method useless. Though we still can fit the data with a correct
shape, the obtained Purcell-factor is absurd and exceeds the
theoretical value. Finally, concerning the third method
(\ref{secfixedpumrate}), the modification of the Purcell factor
induced by the 10\% change in the quality factor amounts to 20 \%,
which is slightly below the stated error-bars due to the imprecision
on the measurement. Therefore, these error-bars are not
significantly increased when allowing the given deviation on the
quality factor. The time resolved measurements also agree within the
error-bars with the CW measurements. The fact that we clearly do not
observe a single exponential decay at resonance, confirms the
hypothesis that other light  sources contribute to the light emitted
into the cavity channel. In particular, for the time resolved
measurement, if not including this light in our model, the lifetime
appears to be pump power dependent, even when we pump way below
saturation which is clearly non-physical. We thus underline that the
commonly adopted criterium that the time resolved spectroscopy of an
exciton has to be made below saturation, might not be sufficient for
CQED experiment. Moreover, if additional emitters are present in the
environment of the considered QD, it might be adequate to include
their presence in the data analysis.

In conclusion, the agreement of the time resolved measurements with
the CW measurements suggests that both methods are reliable. The
dramatic influence of the cavity linewidth uncertainty on the
Purcell-factor error-bars might be a reason for preferring
Q-independent measurements such as time resolved spectroscopy. On
the other hand, the time resolved measurements suffer from a lower
signal-to-noise ratio, and for some systems (photonic crystals, in
particular, where the radiation pattern is less favorable), this
becomes a limiting factor, making the CW measurements more
desirable. In that case, based on above considerations, we advise to
use the method based on the saturation intensity with precaution,
unless a very precise measurement of the cavity quality factor is
available. If this is not the case, the other CW methods here
presented seem more robust against an uncertainty on this parameter.

We thank B. Gayral for fruitful discussions and M. Rigault for his
help and enthusiasm in the initial stages of the measurements, and
we acknowledge QAP for financial support (Contract No. 15848).


\begin{thebibliography}{99}
\bibitem{perfectly} This requires that the emitter-cavity detuning is zero, that the emitter is located at the
field antinode, that it is quasi-monochromatic (linewidth much
smaller than the cavity linewidth), and that the polarisation is the
same as that of the cavity mode.
\end{thebibliography}
\end{document}